\documentclass[twocolumn,superscriptaddress,longbibliography,aps,prb,preprintnumbers,nofootinbib,showkeys]{revtex4-2}

\usepackage[utf8]{inputenc}
\usepackage{graphicx}
\usepackage{bm}
\usepackage{amsmath}
\usepackage{amssymb}
\usepackage{hyperref}
\usepackage{subfigure}
\usepackage{float}
\usepackage{multirow}
\usepackage{booktabs}

\usepackage{varioref}
\usepackage{xr-hyper}
\usepackage[dvipsnames]{xcolor}
\usepackage{nicefrac}
\usepackage{xfrac}
\usepackage{hyperref}
\hypersetup{colorlinks,linkcolor=blue,urlcolor=blue,citecolor=blue}
\usepackage{ulem}
\usepackage{siunitx}
\usepackage{graphicx}
\usepackage{dcolumn}
\usepackage{braket}
\usepackage{wasysym}
\usepackage{textcomp}
\usepackage{orcidlink}
\usepackage{alphabeta}
\usepackage{xcolor}

\begin{document}

\title{Orbital Separation of Charge Order and Superconductivity in La$_{2-x}$Sr$_{x}$CuO$_4$}

\author{I.~Biało~\orcidlink{0000-0003-3431-6102}}
\affiliation{Physik-Institut, Universit\"{a}t Z\"{u}rich, Winterthurerstrasse 190, CH-8057 Z\"{u}rich, Switzerland}

\author{O.~Gerguri}
\affiliation{Physik-Institut, Universit\"{a}t Z\"{u}rich, Winterthurerstrasse 190, CH-8057 Z\"{u}rich, Switzerland}
\affiliation{PSI Center for Neutron and Muon Sciences CNM, 5232 Villigen PSI, Switzerland}

\author{L.~Martinelli~\orcidlink{0000-0003-4978-8006}}
\affiliation{Physik-Institut, Universit\"{a}t Z\"{u}rich, Winterthurerstrasse 190, CH-8057 Z\"{u}rich, Switzerland}

\author{J.~Küspert~\orcidlink{0000-0002-2905-9992}}
\affiliation{Physik-Institut, Universit\"{a}t Z\"{u}rich, Winterthurerstrasse 190, CH-8057 Z\"{u}rich, Switzerland}
\affiliation{European Synchrotron Radiation Facility, 71 Avenue des Martyrs, 38043 Grenoble, France}

\author{J.~Choi~\orcidlink{0000-0003-4616-4345}}
\affiliation{Diamond Light Source, Harwell Campus, Didcot, Oxfordshire OX11 0DE, United Kingdom}
\affiliation{Department of Physics, Korea Advanced Institute of Science \& Technology (KAIST), Daejeon 34141, Republic of Korea}

\author{M.~Garcia-Fernandez~\orcidlink{0000-0002-6982-9066}}
\affiliation{Diamond Light Source, Harwell Campus, Didcot, Oxfordshire OX11 0DE, United Kingdom}

\author{S.~Agrestini~\orcidlink{0000-0002-3625-880X}}
\affiliation{Diamond Light Source, Harwell Campus, Didcot, Oxfordshire OX11 0DE, United Kingdom}

\author{K.~J.~Zhou~\orcidlink{0000-0001-9293-0595}}
\affiliation{Diamond Light Source, Harwell Campus, Didcot, Oxfordshire OX11 0DE, United Kingdom}

\author{E.~Weschke~\orcidlink{0000-0002-2141-0944}}
\affiliation{Helmholtz-Zentrum Berlin für Materialien und Energie, Albert-Einstein-Strasse 15, D-12489 Berlin, Germany}

\author{T.~Kurosawa}
\affiliation{Department of Applied Physics, Hokkaido University, Sapporo 060-8628, Japan}

\author{N.~Momono}
\affiliation{Department of Physics, Hokkaido University, Sapporo 060-0810, Japan}
\affiliation{Department of Applied Sciences, Muroran Institute of Technology, Muroran 050-8585, Japan}

\author{M.~Oda}
\affiliation{Department of Physics, Hokkaido University, Sapporo 060-0810, Japan}

\author{C.~Lin~\orcidlink{0000-0001-8335-2471}}
\affiliation{Physik-Institut, Universit\"{a}t Z\"{u}rich, Winterthurerstrasse 190, CH-8057 Z\"{u}rich, Switzerland}
\affiliation{Stanford Synchrotron Radiation Lightsource, SLAC National Accelerator Laboratory, Menlo Park, USA}

\author{Q.~Wang~\orcidlink{0000-0002-8741-7559}}
\affiliation{Department of Physics, The Chinese University of Hong Kong, Shatin, Hong Kong, China}
\affiliation{State Key Laboratory of Quantum Information Technologies and Materials, The Chinese University of Hong Kong, Shatin, Hong Kong, China}

\author{J.~Chang~\orcidlink{0000-0002-4655-1516}}
\affiliation{Physik-Institut, Universit\"{a}t Z\"{u}rich, Winterthurerstrasse 190, CH-8057 Z\"{u}rich, Switzerland}

\begin{abstract}
We report a combined x-ray absorption and oxygen $K$-edge resonant inelastic x-ray scattering study of charge order in underdoped La$_{2-x}$Sr$_x$CuO$_4$ ($x=0.125$) under uniaxial $c$-axis pressure. We find that compressive $c$-axis strain modifies the charge order only within the superconducting state, with a striking polarization dependence: suppression in the $d_{x^2-y^2}$ channel and enhancement in the $d_{z^2}$ channel. X-ray absorption spectra reveal concomitant strain-induced modifications of the oxygen pre-edge and upper Hubbard band, consistent with increased $d_{z^2}$ orbital admixture. Our results suggest that $c$-axis pressure drives an orbital separation between superconductivity, rooted in $d_{x^2-y^2}$ states, and charge order, which gradually shifts to the $d_{z^2}$ channel. This orbital separation reveals a way for superconductivity and charge order to coexist in the cuprates with minimal competition. Furthermore, it suggests that the multi-order phase diagram of La$_{2-x}$Sr$_{x}$CuO$_4$ cannot be realistically described within single band models usually used to describe cuprate physics.
\end{abstract}

\maketitle

Doped cuprate materials present a complex phase diagram with multiple broken symmetries~\cite{keimer_quantum_2015,ghiringhelli_long-range_2012,WuNat2011,ChangNP2012,robinson_anomalies_2019}.
For hole doped cuprates, charge order is present in virtually all known systems~\cite{kivelson2003How,WuNat2011,ghiringhelli_long-range_2012,ChangNP2012}. This phenomenon is especially stable at and around the specific 1/8 doping. In the La-based cuprates, it couples to spin order to form a stripe structure~\cite{TranquadaNat1995,tranquada_neutron-scattering_1996}. The phase diagram of this electronic order has been established from temperature, doping, and magnetic field dependent studies~\cite{HuckerPRB2011,blanco-canosa_resonant_2014,tabisSynchrotronXrayScattering2017,WuNat2011,Klauss2000}.
In-plane uniaxial pressure experiments have been conducted to clarify the charge ordering structure~\cite{ChoiPRL2022,simutisSingledomainStripeOrder2022}. Large scattering volumes have been recorded to understand the atomic distortions associated to charge ordering~\cite{searsStructureChargeDensity2023,ForganNC2015}. As such, the charge order in cuprates is well characterized.

The interaction between charge order and superconductivity is less well understood. In general, these two electronic orders are believed to compete~\cite{robinson_anomalies_2019,kivelson2003How,tranquada_evidence_1995,comin_charge_2014}. This competition manifests itself through a partial suppression of charge order inside the superconducting state~\cite{ChangNP2012,hucker_competing_2014} and by a weakening of the superconducting temperature and critical field close to $1/8$ doping. It is also the root cause of the magnetic field enhancement of charge order, which occurs exclusively within the superconducting state. Yet, it is still not fully understood how superconductivity and charge order coexist, both spatially and electronically, as new measurements suggest a more complex interaction than a simple competition~\cite{kuspert_engineering_2024, wen_enhanced_2023}.

\begin{figure*}
\centering
\includegraphics[width=\textwidth]{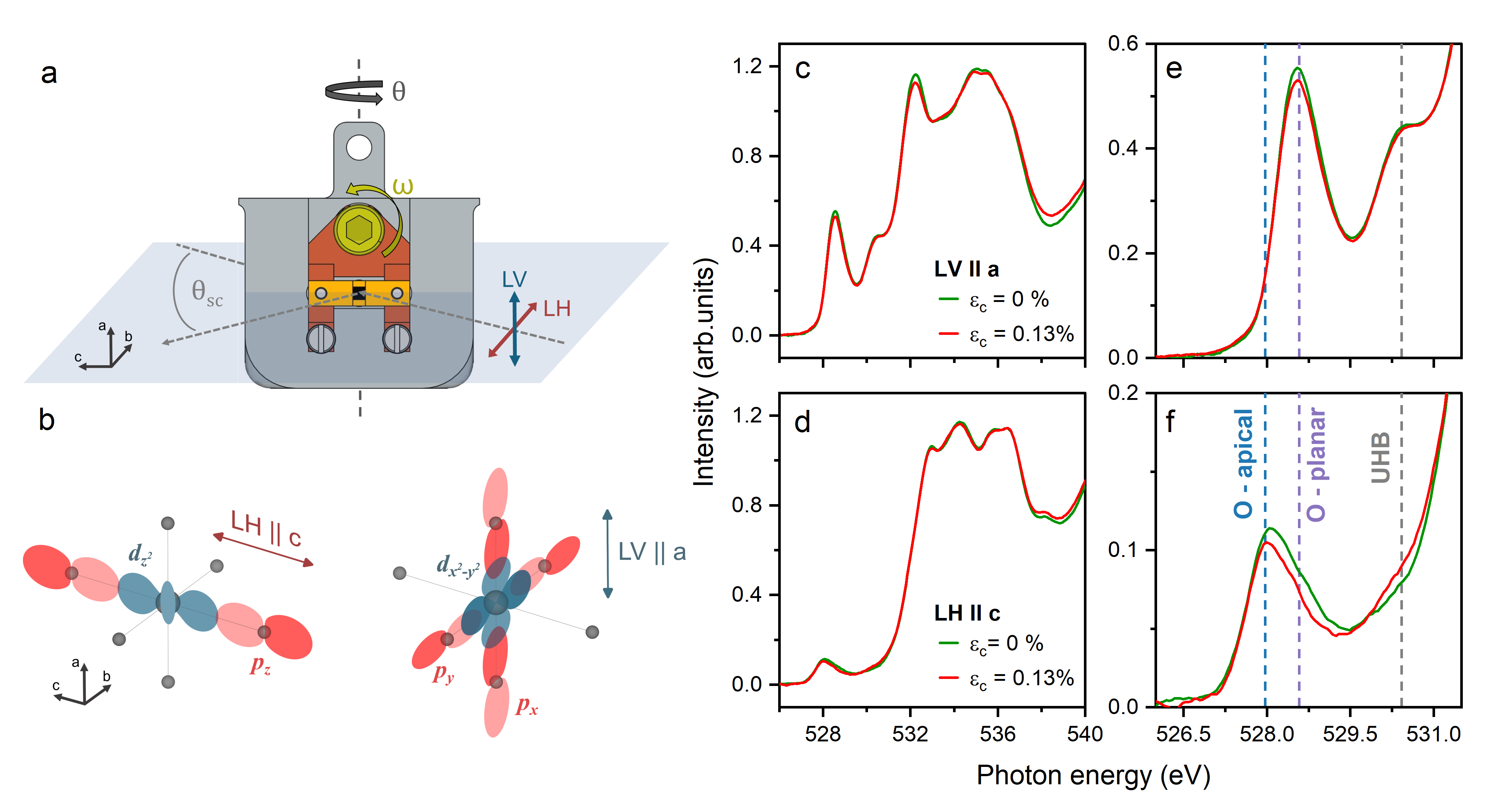}
\caption{(a) Schematic representation of the uniaxial strain device. The sample is mounted with the crystallographic \textit{c}-axis aligned within the scattering plane. The scattering angle $\theta_{SC}$ is controlled by the $\theta$ manipulator rotation angle. Uniaxial strain is tuned by the screw rotation $\omega$. (b) Orientation of \textit{p} oxygen and \textit{d} copper orbitals with respect to the polarization of incident x-ray beam (LH - horizontal, LV - vertical). (c-f) Oxygen $K$-edge x-ray absorption of strained and unstrained LSCO $x=0.125$ at 18~K. (e,f) The oxygen $K$-edge features specific for LSCO~\cite{chen_electronic_1991} are shown in details for LV (e)  and LH polarization (f) and marked with corresponding dashed lines: blue - apical oxygen excitation, violet - planar oxygen, grey - upper Hubbard band.}
\label{Fig1}
\end{figure*}

Uniaxial pressure is an effective tuning parameter to manipulate charge order in superconducting cuprates~\cite{martinelliDecouplingStaticDynamic2025,kim_uniaxial_2018,kuspert_engineering_2024,guguchia_designing_2024,gao_structural_2025}. In particular, $c$-axis pressure experiments have revealed the emergence of a pressure-induced ground state in which the competition between charge order and superconductivity is significantly reduced~\cite{kuspert_engineering_2024}. The length of the $c$-axis lattice parameter itself positively correlates with the superconducting transition temperature $T_c$~\cite{sakakibara_two-orbital_2010,pavarini_band-structure_2001}. Compressive $c$-axis pressure thus reduces the transition temperature~\cite{sato_increase_1997,locquet_epitaxially_1997,nakamura_tc_1999,takeshita_gigantic_2004}. These observations have two causal links. First, a shorter apical oxygen distance effectively produces a smaller anti-ferromagnetic exchange interaction, as it increases the charge-transfer gap and effective Hubbard energy $U$~\cite{weber_scaling_2012}. Assuming that superconductivity is magnetically mediated, it generates a lower $T_c$~\cite{weber_apical_2010}. In this perspective, tuning the $c$-axis lattice parameter induces a magnetic analog to the isotope effect~\cite{ofer_magnetic_2006}. Second, the $c$-axis lattice parameter shapes the crystal field environment and thereby affects the orbital polarization~\cite{SalaNJP2011}. Large $c$-axis distills hole occupation into the $d_{x^2-y^2}$ orbital. In contrast, a smaller $c$-axis parameter enhances the admixture of $d_{z^2}$ orbital character in the hole states. The $d_{z^2}$ "contamination" is unfavorable for superconductivity~\cite{sakakibara_two-orbital_2010,sakakibara_origin_2012}. An outstanding question is how charge order and superconductivity simultaneously react to this orbital contamination.

To address these questions, we designed a resonant inelastic x-ray scattering (RIXS) experiment that allows us to study the charge ordering phenomenon upon the application of $c$-axis pressure. The advantage of resonant scattering techniques is that the incident light polarization offers insight into the orbital character. Consistent with non-resonant diffraction experiments, we show that charge order is influenced by $c$-axis pressure only within the superconducting state. Using linear horizontal (LH) incident light, charge order is enhanced at $T<T_c$. Conversely, it is partially suppressed when probed with linear vertical (LV) polarization. Overall, the suppression in the LV channel is stronger than the enhancement in the LH channel. We discuss these observations in terms of the orbital separation between charge order and superconductivity. 
The implication of our work is that phase competition is incompatible with a single band picture.

\begin{figure*}
\includegraphics{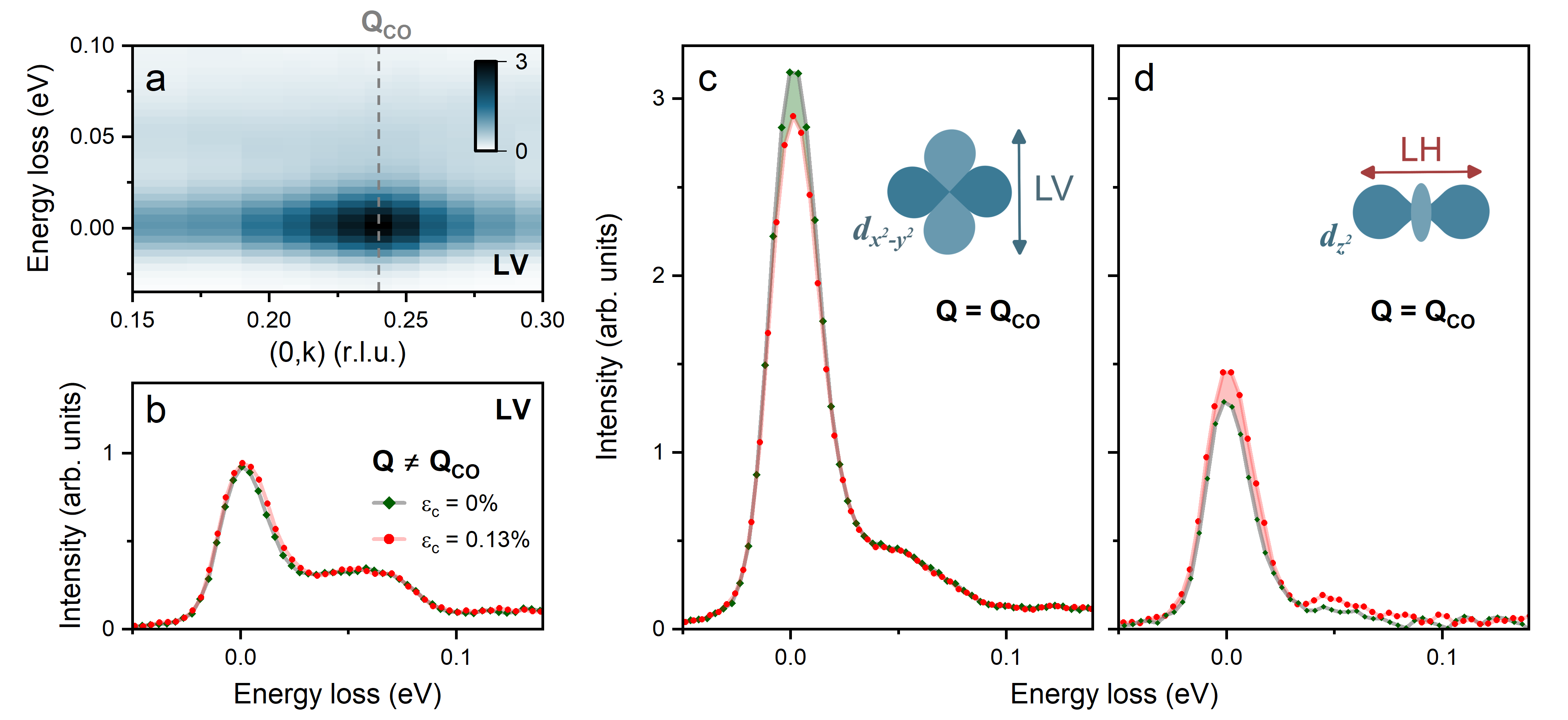}
\caption{RIXS spectra at oxygen K-edge under \textit{c}-axis strain application, at $T<T_c$. (a) RIXS intensity versus in-plane momentum map collected with LV light polarization for unstrained condition. The vertical dashed line marks $Q_{CO}=(0,0.24)$. (b) RIXS spectra collected with LV light polarization, at $Q=(0,0.30)$ -- away from the charge ordering vector. (c-d)  RIXS spectra at $Q_{CO}$. Insets schematically present the polarization of incident beam and the symmetry of the corresponding Cu $d$ orbitals, probed indirectly through hybridization with oxygen. The shaded areas demonstrate the difference within elastic signal induced by \textit{c}-axis strain application.}
\label{Fig2}    
\end{figure*}

\label{Sec:Methods}
High quality single crystals of La$_{2-x}$Sr$_x$CuO$_4$ ($x=0.125$) were grown using the floating zone method~\cite{chang_tuning_2008}. The chosen level of doping optimizes the charge order amplitude~\cite{julien_magnetic_2003}. Tetragonal crystallographic axes were identified using the x-ray Laue method. From the crystal rods, facets normal to the $a,b,$ and $c$ axes were cut. The precision of the $c$-axis facet was within $2^{\circ}$. The facet spanned by the $a$ and $c$ axes was polished using a combination of silicon carbide (SiC) and sapphire lapping disks with roughness down to 0.5~$\mu$m, attaining a mirror-like surface (see End Matter).

Our x-ray absorption spectroscopy (XAS) and RIXS experiments were carried out at the I21 beamline~\cite{zhou_i21_2022} at the Diamond Light Source. For this beamline, a uniaxial pressure cell --- capable of applying both compressive and tensile strain~\cite{lin2024Uniaxial} --- was adapted~\cite{martinelliDecouplingStaticDynamic2025}. Its design and a sketch of experimental geometry are shown in Fig.~\ref{Fig1}a. The sample (black) is stabilized between the cell's jaws (orange) by an epoxy resin. The relationship between screw rotation and crystal lattice deformation is included in End Matter. In this work, we focused on the application of compressive strain along the crystallographic $c$-axis. The crystal was mounted so that linear vertical (LV) light polarization, perpendicular to the scattering plane, is aligned with the $a$-axis, while linear horizontal (LH) polarization is aligned with the $c$-axis (Fig.~\ref{Fig1}a,b). We performed measurements at the oxygen $K$-edge, with a combined energy resolution of 25~meV (high-resolution mode) or 40~meV (medium resolution mode). The scattering angle was fixed at $2\theta^{\circ}=154$. By rotating the incident angle $\theta$ (see Fig.~\ref{Fig1}a), RIXS spectra were recorded along the reciprocal $b$-axis direction, allowing access to the $(0,k,\ell)$ scattering plane. The intensity of RIXS spectra is normalized to the spectral weight of $dd$ charge transfer excitations, as in Ref.~\cite{GhiringhelliSci2012,Wang2021,arpaia2023Signaturea,LinPRL2020}.

Fig.~\ref{Fig1}(c-f) displays oxygen XAS of LSCO $x=0.125$ collected with different incident light polarizations. As the crystal field environment around the apical and planar oxygen is different, the pre-edge has two resonances split by about 0.6~eV. The lowest energy resonance is associated with the apical oxygen~\cite{fatuzzo_spin-orbit-induced_2015,moretti_sala_orbital_2014}. As previously demonstrated, the planar resonance is strongest when light polarization enhances the $p_x$ cross section. By contrast, the apical resonance is enhanced when the $p_z$ cross section is optimized. The resonance associated with the upper Hubbard band (UHB) is found at about 2~eV above these pre-edge features~\cite{brookes_stability_2015}. Again, consistent with previous reports~\cite{chen_out--plane_1992}, this resonance appears strongest in the $p_x$ channel, suggesting a dominant $d_{x^2-y^2}$~character.

Upon application of the $c$-axis strain, the resonances are modified. The pre-edge peak amplitude is reduced upon compressive $c$-axis pressure: the planar resonance is diminished irrespective of light polarization, while the apical resonance is reduced when probing through the $p_z$ channel. A small energy shift of both resonance energies is also detected (Fig.~\ref{Fig1}f), indicating that the $c$-axis pressure modifies the crystal field environment. Likewise, the UHB resonance reacts to $c$-axis pressure. A very small reduction is found when probing through the $p_x$ channel, whereas a substantial increase is found through the $p_z$ orbital.

The charge order in underdoped LSCO manifests itself through diffraction peaks at $Q=\tau+Q_{co}$, where $\tau$ are fundamental lattice Bragg reflections, and $Q_{co}=(\pm\delta,0,\pm0.5)$ and $(0,\pm\delta,\pm0.5)$  with $\delta \approx 1/4$~\cite{Christensen14}. 
We performed RIXS studies to intersect the charge order along the $(0,k,\ell)$ direction at the planar oxygen resonance (at about 528.6~eV).
Given the two-dimensional nature of the charge order~\cite{croft_charge_2014}, 
the change in $\ell$ throughout the rocking scans does not affect the results.

\begin{figure*}
\includegraphics{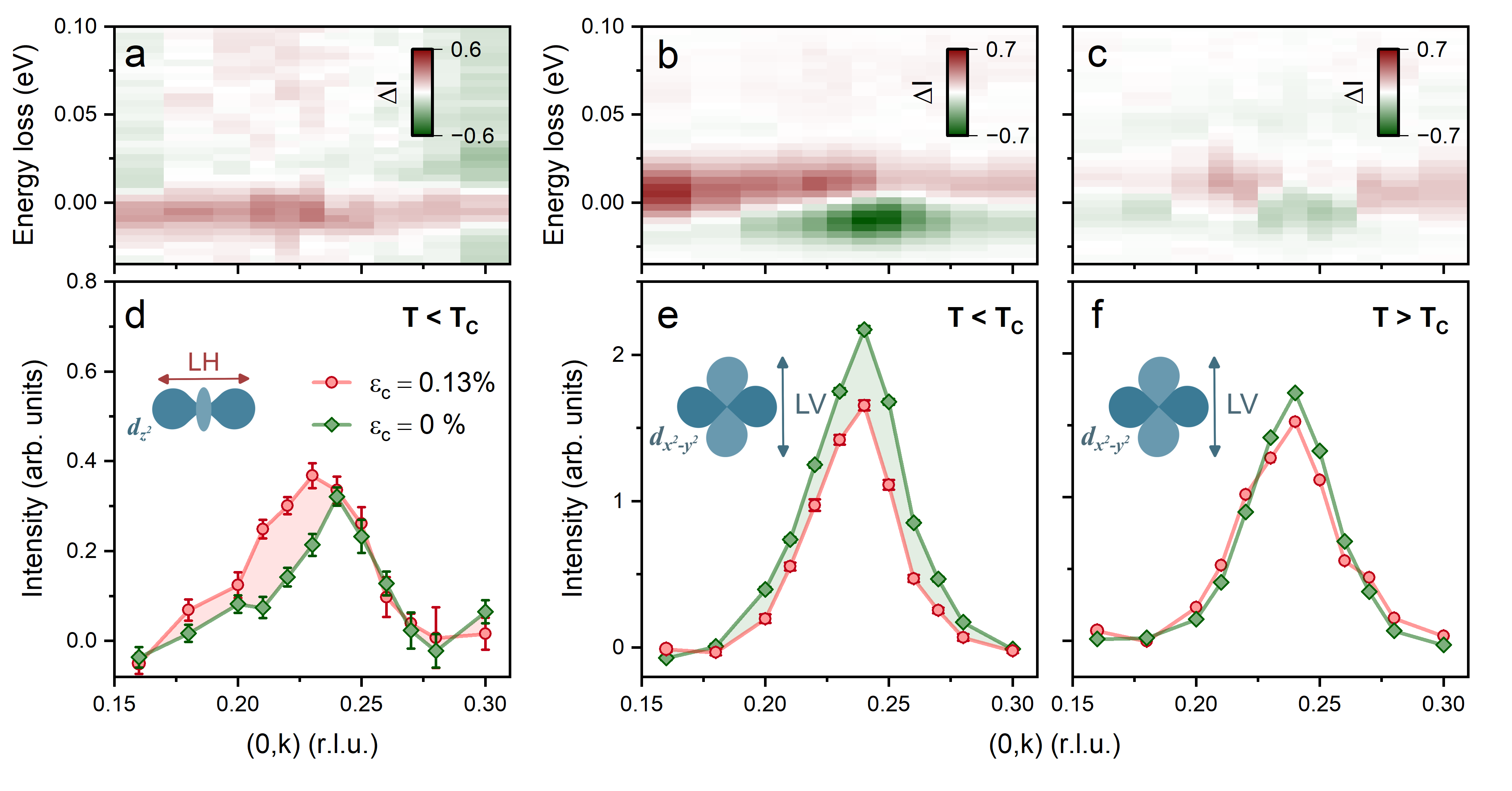}
\caption{(a-c) Difference in RIXS intensity ($\Delta I$) between spectra taken with and without applied maximum strain (0.13\%). Temperature and polarization conditions are indicated in the corresponding panels below. (d-f) Elastic $Q$-scans through the charge ordering vector. Intensity of charge order for unstrained (green) and maximally strained sample (red) measured below (18~K) and above $T_c$ (40~K). Insets schematically represent the polarization of incident beam and the symmetry of probed copper orbitals.} 
\label{Fig3}
\end{figure*}
The RIXS momentum map along the $(0,k)$ direction is presented in Fig.~\ref{Fig2}a. We explored the influence of $c$-axis strain on the charge order while probing with LV/LH polarized light. In Fig.~\ref{Fig2}(b-d), we show RIXS spectra at and away from the $Q_{co}$. Measurements were performed below $T_c$, where the competitive nature of charge order and superconductivity has been reported~\cite{ChangNP2012, kuspert_engineering_2024}. Elastic scattering not associated with the charge order reflection displays no appreciable uniaxial $c$-axis pressure effects (see Fig.~\ref{Fig2}b). By contrast at $Q_{co}$, a $c$-axis pressure effect is observed. The charge order reflection is respectively enhanced and reduced, dependent on the light polarization.  This is the key observation of this study. It can be analyzed using two different methodologies. 

Qualitatively, the change in RIXS intensity under applied pressure can be presented in the form of differential momentum maps of $\Delta I=I(\varepsilon_c=0.13\%)-I(\varepsilon_c=0\%)$. We detect an \textit{increase} in the CO intensity with applied $c$-axis strain when probing at the planar oxygen resonance with LH polarized light (Fig.~\ref{Fig3}(a)). The effect is opposite for LV polarization (Fig.~\ref{Fig3}(b)). Importantly, this pressure effect effectively vanishes above $T_c$, as shown in Fig.~\ref{Fig3}(c). To quantify the change in charge order intensity under applied strain, we extracted the elastic scattering through fitting (described in detail in the Supplementary Materials). The influence of $c$-axis pressure on the charge order reflection is presented in Fig.~\ref{Fig3}(d-f). The intensity of the charge order peak decreases in the LV ($d_{x^2-y^2}$) channel but increases in the LH ($d_{z^2}$) channel. In the normal state, the intensity of the charge order reflection remains constant. 

\begin{figure*}[]
\includegraphics{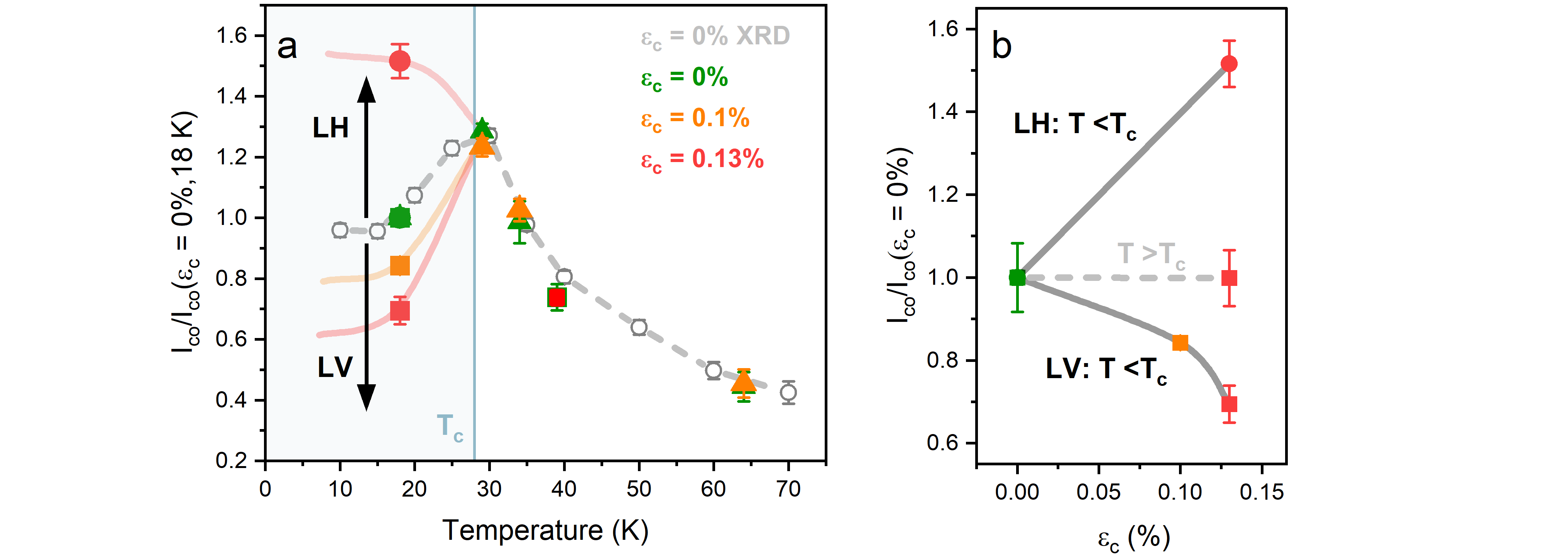}
\caption{Temperature dependence of charge order under uniaxial $c$-axis strain. (a) Integrated intensity of charge order ($I_{co}$) under uniaxial strain as a function of temperature, overlaid with unstrained x-ray diffraction (XRD) data adapted from Ref.~\cite{kuspert_engineering_2024}. Filled squares (triangles) denote high-resolution (medium-resolution) RIXS measurements performed with LV-polarized light, while filled circles correspond to high-resolution data collected with LH polarization. Black arrows schematically indicate the evolution of charge order under strain, depending on the incident light polarization. Data are normalized to unstrained values at 18~K. (b) Integrated intensity of charge order as a function of strain, measured with different light polarizations, both within and outside the superconducting state, as indicated. Data are normalized to the corresponding unstrained value.}
\label{Fig4}
\end{figure*}

The temperature dependence of the strain effect is summarized in Fig.~\ref{Fig4}a. In the absence of applied strain, the intensity of charge order follows the characteristic temperature dependence widely reported for underdoped LSCO~\cite{croft_charge_2014,wen_observation_2019,vonarxFateChargeOrder2023,kuspert_engineering_2024} and remains unaffected by the strain. Below $T_c$, 
strain affects the charge order amplitude, as reported in Fig.~\ref{Fig4}b.

We report significant pressure-induced changes in both the XAS oxygen pre-edges and the upper Hubbard band. Compressive strain along the $c$-axis shortens the apical oxygen distance, reducing the tetragonal-split between $d_{z^2}$ and $d_{x^2-y^2}$ states~\cite{khomskii2014transition}, which in turn promotes hybridization between the two bands. This provides an explanation for the observed decrease in XAS intensity at the planar oxygen site probed through the $p_{x}$ orbital. Interestingly, the apical oxygen site probed via the $p_{z}$ orbital also exhibits a reduction, suggesting that $c$-axis compression not only shortens the lattice parameter but also induces or enhances orthorhombic distortions, which would reduce the $d_{z^2}$–$p_{z}$ orbital overlap. In contrast, the UHB shows a comparatively smaller response to $c$-axis pressure. Notably, the UHB resonance is weakened in the LV channel while strengthened in the LH channel, which may indicate an increase in $d_{z^2}$ hole population within the UHB at the expense of $d_{x^2-y^2}$ holes.

\begin{figure}[]
\includegraphics{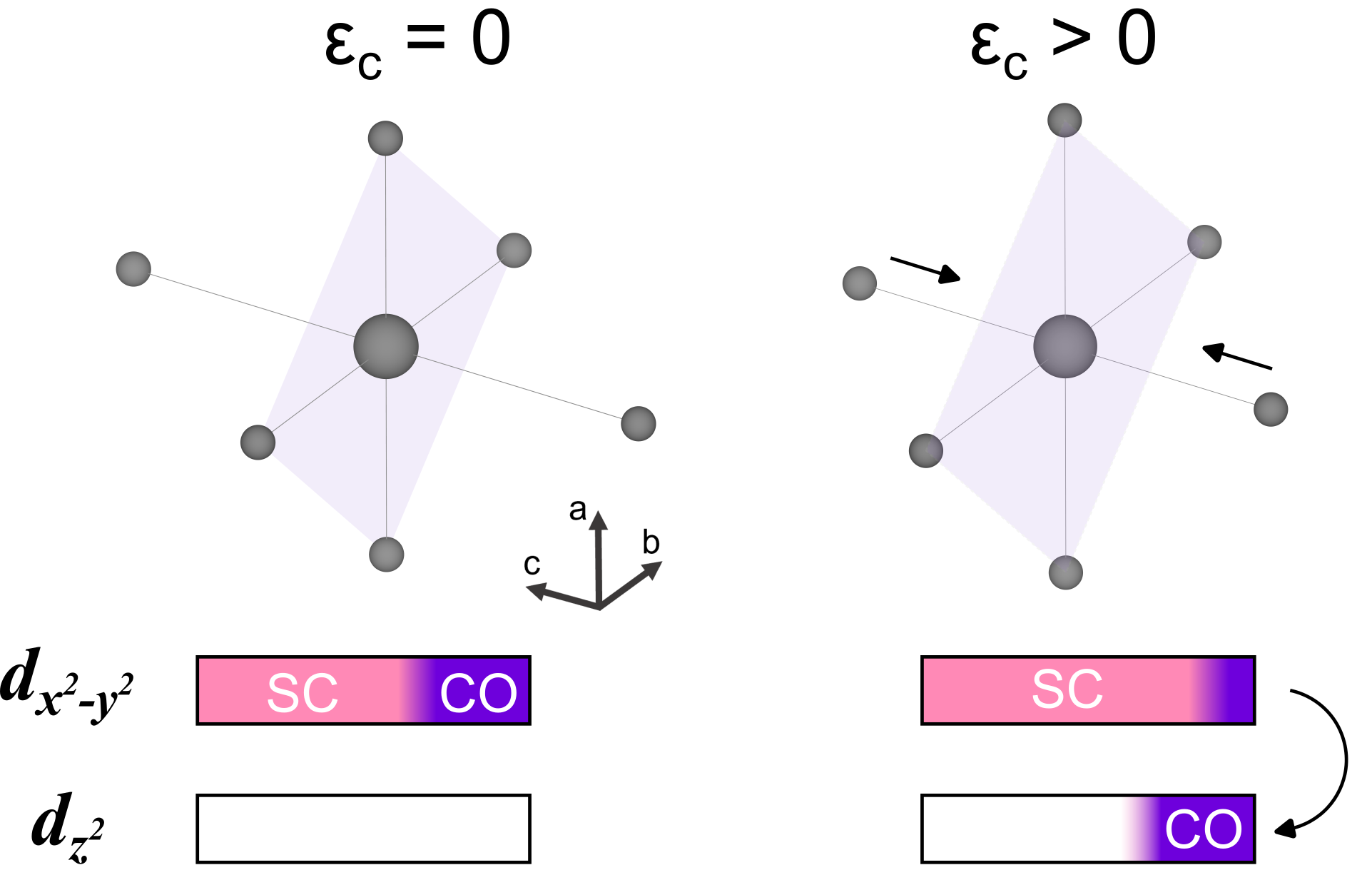}
\caption{Orbital separation of superconductivity and charge order with uniaxial strain. \textit{c}-axis strain deforms copper - oxygen octahedra, reducing the energy split between $d_{x^2-y^2}$ and $d_{z^2}$ orbitals. As a result, the carriers involved in charge order (CO) formation shift to $d_{z^2}$ states to avoid competition with superconductivity (SC).}
\label{Fig5}
\end{figure}

Uniaxial strain also influences the charge ordered state. Previously, non-resonant diffraction demonstrated that charge order is enhanced by $c$-axis pressure within the superconducting state~\cite{kuspert_engineering_2024}. This  indicates that strain tunes the competition between charge order and superconductivity. Here, we confirm that the $c$-axis pressure effect on charge order emerges only within the superconducting state. However, by using polarized light, our resonant experiment uncovers an important new aspect. The charge order is only enhanced in the LH channel, which predominantly probes the $d_{z^2}$ orbital. In contrast, in the $d_{x^2-y^2}$ orbital, the charge order reflection is suppressed. 

In cuprates, superconductivity is promoted in $d_{x^2-y^2}$ orbitals hybridized with $p_{x,y}$ states~\cite{sakakibara_origin_2012}. The pairing strength within this channel is, however, highly influenced by the presence of $d_{z^2}$ states near the Fermi level~\cite{sakakibara_two-orbital_2010, matt_direct_2018}. By applying compressive $c$-axis pressure, we shorten the apical oxygen distance and induce more $d_{z^2}$~states, suppressing superconductivity~\cite{SalaNJP2011,sakakibara_origin_2012}. 
The presence of two orbital channels ($d_{x^2-y^2}$ and $d_{z^2}$) opens new possibilities 
for the co-existence of superconductivity and charge order. It is clear that superconductivity 
involves only 
the $d_{x^2-y^2}$ channel~\cite{sakakibara_two-orbital_2010,sakakibara_origin_2012}. Charge order, however, can emerge within both channels~\cite{gaina_long-ranged_2021}. A shift of the doped holes that gives rise to the charge order modulation to the $d_{z^2}$ states would provide a pathway for reduced competition within the $d_{x^2-y^2}$ channel (Fig.~\ref{Fig5}).
We are thus speculating on a gradual orbital separation between superconductivity and charge order. This would explain why, in the non-resonant experiments, the magnetic field effects on the charge order decline with uniaxial $c$-axis pressure~\cite{kuspert_engineering_2024}. As the orbital separation evolves, the direct phase competition is avoided, and with it, the magnetic field effects on the charge order inside the superconducting state vanish.

The orbital scenario is supported by quantitative arguments.
From photoemission experiments, it is known that the $d_{z^2}$ hybridization in LSCO is weak~\cite{matt_direct_2018}. In fact, it is estimated that near the Fermi level, the states are 99\% of $d_{x^2-y^2}$~character. 
Now, the question is how many electrons are involved in the charge ordering. This can be estimated from non-resonant diffraction. The ratio between the intensity of charge order and neighboring Bragg reflections scales with the number of electrons involved in the modulation. For YBa$_2$Cu$_3$O$_{7-\delta}$, charge order is produced by a modulation of about 1\% of the available electrons~\cite{ChangNP2012}. 
Quantitative comparisons indicate that LSCO’s charge order is approximately an order of magnitude weaker than the charge order in YBa$_2$Cu$_3$O$_{7-\delta}$~\cite{miao_charge_2021,Thampy2013} or the well-developed stripe order in n La$_{2-x}$Ba$_x$CuO$_4$~\cite{Christensen14}.
Thus, it is likely that the charge order in LSCO is composed of less than 1\% of the available electrons. It is therefore not unreasonable to think that the small fraction of $d_{z^2}$ states can have a significant effect on the charge ordering.


In summary, we have carried out a uniaxial $c$-axis pressure experiment on the charge order in LSCO with $x=0.125$. Upon the application of $c$-axis pressure, we demonstrate the tunability of the charge order parameter. This $c$-axis strain effect occurs only within the superconducting state. Both suppression and enhancement of charge order are observed, depending on the incident light polarization. We interpret this as an orbital separation effect. While superconductivity emerges from $d_{x^2-y^2}$ states; the charge order manifests in both $d_{x^2-y^2}$ and $d_{z^2}$ channels. To avoid competition for $d_{x^2-y^2}$ states, charge order shifts to the $d_{z^2}$ states within the superconducting state. $c$-axis pressure promotes $d_{z^2}$ states near the Fermi level, which in turn enables an orbital decoupling of superconductivity and charge order.   




\vspace{0.5cm}
\textit{Acknowledgements} -- I.B. and L.M. acknowledge support from the Swiss Government Excellence Scholarship under project numbers ESKAS-Nr: 2022.0001 and ESKAS-Nr: 2023.0052.
I.B. and L.M. also acknowledge support from UZH Postdoc Grants, project numbers FK-23-113 and FK-23-128.
L.M acknowledges support from the Swiss National Science Foundation under Spark project CRSK-2\textunderscore220797.
Q.W. is supported by the Research Grants Council of Hong Kong (ECS No. 24306223), and the Guangdong Provincial Quantum Science Strategic Initiative (GDZX2401012).
J. Choi acknowledges financial support from the National Research Foundation of Korea (NRF) funded by the Korean government (MSIT) through Sejong Science Fellowship (Grant No. RS-2023-00252768). 
J.K., L.M., and J.C. acknowledge support from the Swiss National Science Foundation under grant number 200021-188564. J.K. is further supported by the PhD fellowship from the German Academic Scholarship Foundation. We acknowledge Diamond Light Source for time on Beamline I21 under Proposal MM34663. 
We acknowledge BESSY II for providing beam time on beamline UE46-PGM1 under Proposal No. 232-12330-ST-1.1-P.


\bibliography{ZoteroIB,LQMR,lsco_ref}

\begin{figure}
    \centering
    \includegraphics{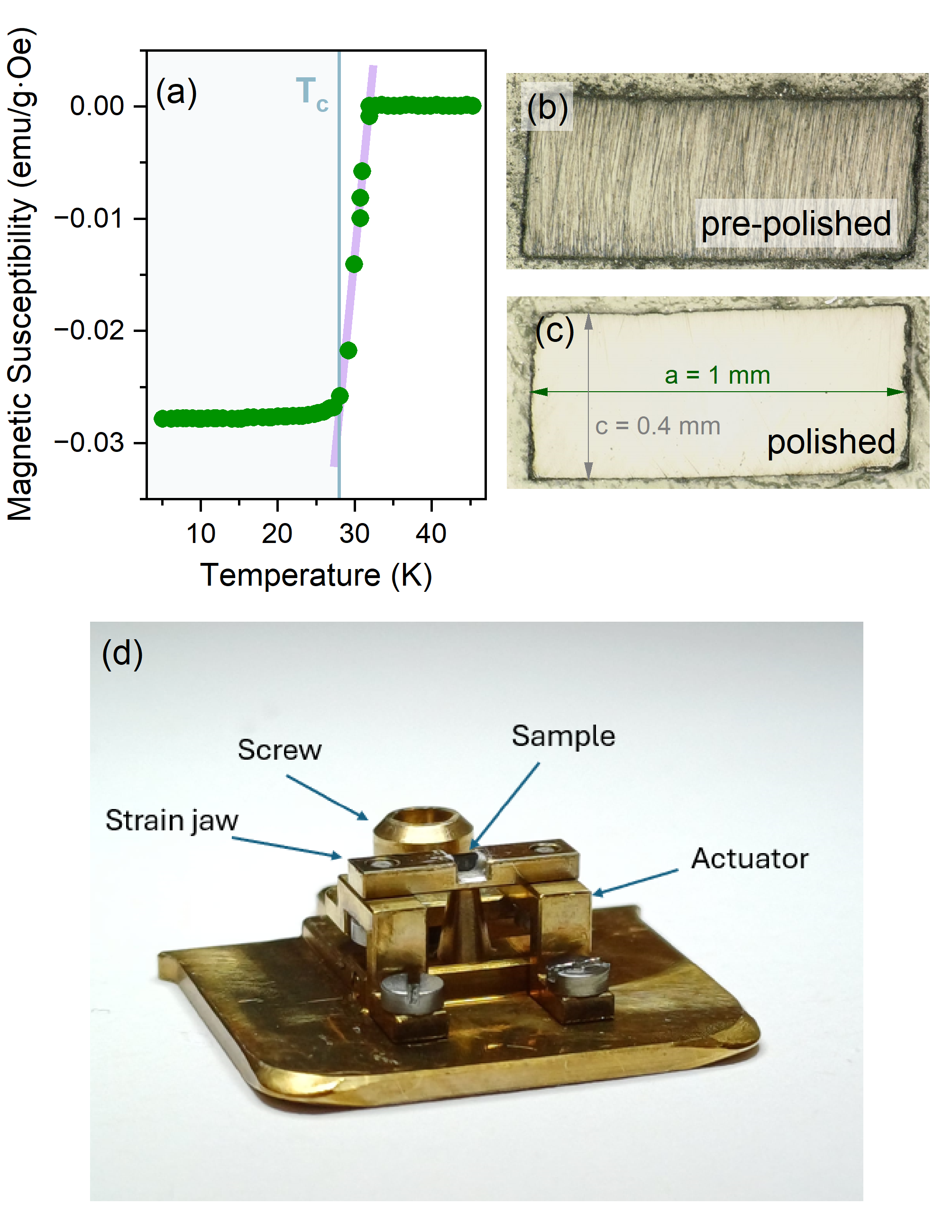}
    \caption{(a) Magnetic susceptibility of La$_{2-x}$Sr$_x$CuO$_4$ ($x=0.125$) with $T_c$ as indicated. (b) Pre-polished sample surface. (c) The quality of the sample surface after the polishing. The sample is naturally black but the photo is taken under reflecting light. (d) Uniaxial strain device with the LSCO sample inserted. Compressive strain was applied by a counter-clockwise rotation of the driving screw. }
    \label{FigS1S2}
\end{figure}

\onecolumngrid 
\vspace{0.2cm}
\begin{center}
    \textbf{End Matter} 
\end{center}
\twocolumngrid 
\textit{Appendix A: LSCO crystals} -- 
The critical temperatur, $T_c$, was estimated to be $28$~K by magnetization measurements (Fig.~\ref{FigS1S2}(a)). With Laue diffraction, samples were aligned and cut from the rod into the shape of cubes. The size $0.4\times0.7\times1$~mm along the $a$, $b$ and $c$ axes, respectively, was achieved. The longest dimension $c\sim1$~mm coincides with the direction of strain application. The \textit{ac-plane} of the crystal was polished using SiC polishing paper and sapphire lapping disks, with a roughness down to 0.5 $\mu$m see Fig.~\ref{FigS1S2}(b,c). This way, we reached a mirror like sample surface that minimizes the elastic contribution in resonant inelastic x-ray scattering spectra.

\textit{Appendix B: Uniaxial strain device} -- Our strain device (photo shown in Fig.~\ref{FigS1S2}(d)) for resonant inelastic x-ray scattering experiments is described in~\cite{martinelliDecouplingStaticDynamic2025}. It is constructed on a base of a standard Omicron plate. The sample is mounted in the groove of a T-shaped element (the strain jaw), which lies on the base of the device and on the actuator element. The movement of the driving screw bends the actuator up or down depending on the sense of rotation. The strain jaw is then deformed, and the gap either shrinks, applying compressive strain, or opens, applying tensile strain. In particular, when the driving screw is turned anti-clockwise, compressive strain is applied. The application of strain is ensured by a circlip mounted on the driving screw, below the actuator. The plate, platform, actuator, and driving screw are made of BeCu, while the circlip is made of stainless steel. 

\begin{figure}[h]
    \centering
    \includegraphics{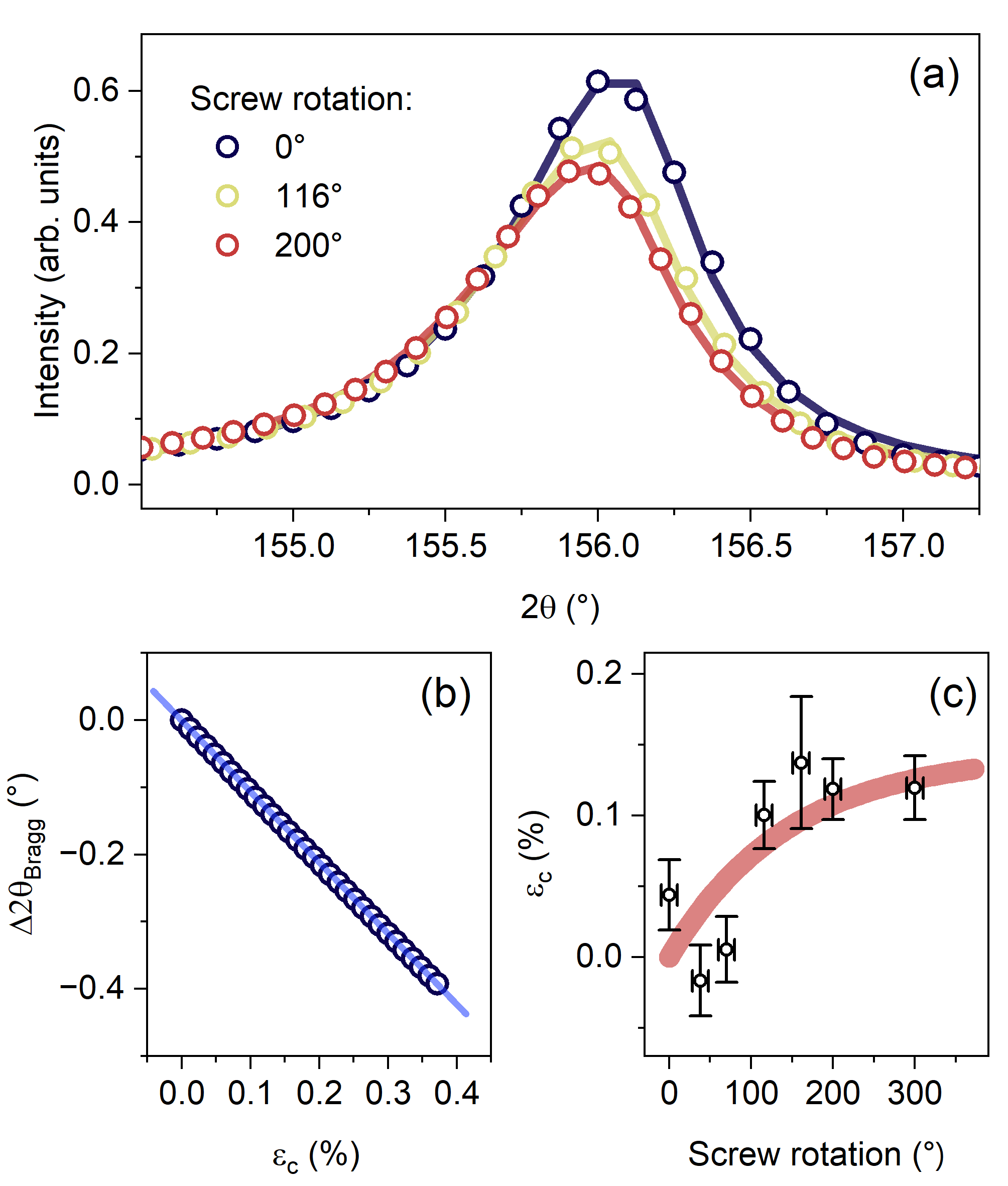}
    \caption{The (-101) Bragg reflection measured on a LSCO sample as a function of applied strain, quantified in terms of driving screw rotation angle. Points are the result of a $\theta-2\theta$ scan. Solid lines represent Lorentzian fits. (b) $c$-axis strain calculated based on the shift of the Bragg angle shift. The solid line is a linear fit to the simulated points. (c) Strain values extracted from the Bragg angle shift presented as a function of driving screw rotation angle.}
    \label{FigS3}
\end{figure}

\textit{Appendix C: Strain calibration} -- We estimated the applied uniaxial strain using soft x-ray diffraction at the UE46-PGM1 beamline of BESSY~II. Considering the scattering geometry with the $c$-axis lying in the scattering plane, we tracked the angular position of the $(-101)$ Bragg reflection as a function of the screw rotation. Fig.~\ref{FigS3}(a) shows representative $\theta$--$2\theta$ scans collected at a photon energy of 1750~eV. The scans were fitted using an asymmetric Lorentzian profile (solid lines), from which we extracted the Bragg angle. As the screw is rotated, the Bragg peak systematically shifts to lower angles.

To convert the measured angular shift, $\Delta2\theta_{\mathrm{Bragg}}$, into the corresponding lattice deformation, we simulated the change in lattice parameters associated with the observed peak movement, as shown in Fig.~\ref{FigS3}(b). We assumed a Poisson ratio of $\nu_{bc} \simeq 0.35$ relating the in-plane and out-of-plane lattice responses, consistent with earlier $c$-axis strain studies on La$_{2-x}$Sr$_x$CuO$_4$~\cite{ChoiPRL2022,meyerStrainrelaxationCriticalThickness2015}. This yields a linear relationship between the $c$-axis strain, $\varepsilon_c$, and the Bragg-angle shift. Using this relation, we constructed a calibration curve for our uniaxial strain device, shown as the red line in Fig.~\ref{FigS3}(c). The curve accounts for the direction of applied strain and the specific sample geometry (size and shape).

\end{document}